\documentclass[11pt,leqno]{article}

\usepackage{amsmath}
\usepackage{amssymb}
\usepackage{amsthm}
\usepackage{array}
\usepackage{authblk}
\usepackage[backend=bibtex]{biblatex}
\usepackage{listings}
\usepackage{bm}
\usepackage{caption}
\usepackage[utf8]{inputenc}
\usepackage[threshold=0]{csquotes}
\usepackage{enumitem}
\usepackage{float}
\usepackage{fullpage}
\usepackage[top=1in, bottom=1in, left=1in, right=1in]{geometry}
\usepackage{tikz}
\usepackage{graphicx}
\usepackage{mathtools,etoolbox}
\usepackage{setspace}
\usepackage{subcaption}
\usepackage{url}

\usetikzlibrary{shapes.arrows,chains}

\newcommand{\mydef}[1]{\textbf{#1}}

\title{Comparative Analysis of Classic Garbage-Collection Algorithms for a Lisp-like Language}

\author{Tyler Hannan}
\author{Chester Holtz}
\author{Jonathan Liao}

\affil{Department of Computer Science\\University of Rochester\\Rochester, NY 14627, USA}

\date{April 20, 2015}

\begin{document}
\sloppy

{
\singlespacing
\maketitle
}

\begin{abstract}
\noindent In this paper, we demonstrate the effectiveness of Cheney's Copy Algorithm for a Lisp-like system and experimentally show the infeasability of developing an optimal garbage collector for general use. We summarize and compare several garbage-collection algorithms including Cheney's Algorithm, the canonical Mark and Sweep Algorithm, and Knuth's Classical Lisp 2 Algorithm. We implement and analyze these three algorithms in the context of a custom MicroLisp environment. We conclude and present the core considerations behind the development of a garbage collector---specifically for Lisp---and make an attempt to investigate these issues in depth. We also discuss experimental results that imply the effectiveness of Cheney's algorithm over Mark-Sweep for Lisp-like languages.
\end{abstract}

\section{Introduction}

Virtually every modern language has some form of automated garbage collection and memory management. When an object is no longer referenced by a program, it is necessary for the heap space in a machine's memory to be recycled, else one risks a program using up an entire machine's memory. If there is no automatic reclamation of memory, it is the job of the programmer to manually manage his or her memory. Garbage collectors are far from perfect, as discussed and rigorously stated in our analysis. It is commonly understood that garbage collection will always be inefficient and slower in terms of speed when compared with optimal manual memory management, but offer convenience and stability in return.  In this paper, we offer an analysis of three garbage collection systems: Cheney's Copy Collector \cite{hanappe}, The Traditional Mark-Sweep method \cite{fenichel}, and Knuth's Lisp 2 Algorithm \cite{knuth3} and comment on their respective advantages and disadvantages.

\subsection{Background and related work}
As discussed earlier, the convenience afforded by using a garbage collector is an advantage to the programmer. Garbage collectors shift the burden of memory allocation from the programmer to the machine. As a result of the effect garbage collection has on the efficiency and security of the program, automated garbage collection is a heavily researched area of languages and compiler theory, topics of computer science. We analyzed classical works such as Knuth’s Art of Computer Programming Volume 1 and 3 \cite{knuth3}, and John McCarthy’s Essays on garbage collection \cite{mccarthy} to understand the fundamental principles that define an effective automated memory management system. In particular, modern research in garbage collection delves into topics such as virtualization, parallelization, efficiency, or high level cooperation with the operating system. As we have stated, one goal of this paper is to discuss the possibility that developing a general use garbage collector is infeasible and that it is necessary to manufacture different memory management systems depending on the purpose of the program they will be used in.

\subsection{Definitions and terminology}
\textbf{Interpreter}\\
Before developing the garbage collector, we first implemented an interpreter for a lisp-like language - namely following the Micro-Lisp framework specified by John McCarthy\cite{mccarthy}. We will first describe our implementation and the various terms and aspects involved in the structure of our Micro-Lisp. In particular, \mydef{symbolic expressions} are defined as being members of one of two classes: \mydef{lists}, and \mydef{atoms}. Atoms, by definition, are collections of letters, digits or other characters not otherwise defined in the micro-lisp language. Lists, on the other hand, consist of a left parenthesis followed by a head - or CAR -  and a tail - a CDR. Lists end with a closing parenthesis. We say that Micro-Lisp is defined by a set of rules or functions which take an input set and produce some output. Specifically, the functions for which a valid definition exists in our language are given below. Based on convention, we denote \bm{$e$} and \bm{$a$} as expressions, the letter \bm{$v$} as an atom serving as a variable, and \bm{$f$} as a function name.
\begin{enumerate}
  \item QUOTE - The value of (QUOTE $A$) is $A$
  \item CAR - The value of (CAR $e$) is the first element of $e$ where $e$ is defined as a non empty list - i.e. (CAR (QUOTE ($A$ $B$))) returns $A$. 
  \item CDR - The value of (CDR $e$) is the remaining elements of e when the CAR of e is removed where e defined as a non empty list - i.e. (CDR (QUOTE ($A$ $B$))) returns $B$. 
  \item CONS - The value (CONS $e_1$ $e_2$), is the list that results from prefixing value $e_1$ onto the list value $e_2$. Thus value (CONS (QUOTE $A$) (QUOTE $B$))) returns the list ($A$ $B$).
  \item EQUAL - The value of (EQUAL $e_1$ $e_2$) is true if $e_1$ = $e_2$ and false if otherwise.
  \item ATOM - The value of (ATOM $e_1$) is true if $e_1$ is an atom and false if otherwise.
  \item COND - The value of (COND($e_1$ $e_1$) ... ($p_n$ $e_n$)) is the value of $e_i$, where $p_i$ is the the 
first of the $p's$ whose value is not NIL.
  \item DEFINE - The DEFINE function maps a variable to an arbitrary expression.
  \item LAMBDA - Lambda is a construct in the Micro-Lisp language which allows for the definition of anonymous functions.
  \item Operators - We also provide definitions to traditional boolean ( $=$, $>$, $\geq$, $<$, $\leq$...) and arithmetic operators ($+$, $-$, $*$, $/$...)
\end{enumerate}
\textbf{Garbage Collector}\\
Following interpretation, the garbage collector attempts to reclaim \mydef{garbage} as a form of memory management. Garbage refers to memory that is occupied by \mydef{objects} no longer used by the program. Objects refer to a location in memory having a value, such as a variable, data structure, or function. There are multiple different algorithms for garbage collection, but the basic principles consist of finding objects which can no longer be accessed, and then reclaiming the resources that are used by these objects. For our analysis, Mark-Sweep is a type of \mydef{Tracing garbage collection} which determines which objects in the heap to deallocate based upon their reachability from root nodes. Cheney's Algorithm is a \mydef{Copy} or \mydef{Semispace} algorithm which defines two equally sized spaces in memory and moves objects between the two based upon an object's accessibility. The Lisp 2 algorithm is a classical algorithm proposed by Knuth with elements of tracing collection.\\

\noindent
\textbf{Heap}\\
The \mydef{heap} is an abstract structure in memory which contains memory that can be dynamically allocated - in contrast to the \mydef{stack} where memory that can only be allocated in a well-defined order resides. For our definition, the heap is made up of a set of objects, and a particular item known as the \mydef{root}. 

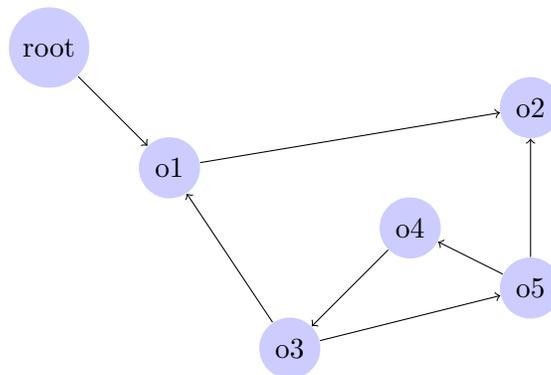
\begin{figure}[H]
	\centering
		\begin{tikzpicture}
          [scale=.8,auto=left, every node/.style={circle,fill=blue!20}]
          \node (r) at (1,10) {root};
          \node (o1) at (3,8)  {o1};
          \node (o2) at (9,9)  {o2};
          \node (o3) at (5,5)  {o3};
          \node (o4) at (7,7)  {o4};
          \node (o5) at (9,6)  {o5};

          \foreach \from/\to in {r/o1,o1/o2,o5/o2,o5/o4,o4/o3,o3/o5,o3/o1}
            \draw [->] (\from) -- (\to);
        \end{tikzpicture}
	\caption{Graph Representation of Heap}
	\label{fig:heapex}
\end{figure}

In Figure~\ref{fig:heapex}, we present an example of the heap represented as a directed graph of a root and object nodes represents its size. In this graph, the nodes are represented as objects $o=(a,s)$ - for a representing the pointer to the head of the object and s representative of the object's size - while the edges between objects are the references between objects. We say that each object has an address, size, and a set of references to other objects. It is important to note that no two objects may overlap in memory, or that an arbitrary object must be defined in memory entirely before or after any other object. Rigorously, we says that given two objects in the heap $o_1 = (a_1, s_1), o_2 = (a_2, s_2)$, $a_1+s_1 \leq a_2$ or $a_2+s_2 \leq a_1$. It then follows that we can preform some operations on the heap. Common operations to keep the heap's size in check such as removing nodes with no path to the root, or coalescing free-space nodes that are adjacent in memory are traditionally effective, but tax the system.
    
\section{Algorithms}
We examine three examples of classical garbage collection algorithms. Since there are many variations to the following versions, an attempt has been made to primarily analyze and compare only the algorithms defined in their original form.  

\subsection{Tri-color marking}
A common abstraction implemented by modern tracing collectors is the \mydef{tri-color marking scheme}\cite{wiki:garbage_collection}. By definition, objects are assigned a color - white, black and gray. We say that the
\textbf{white} set is the set of objects that are candidates for reaping. This set is also known as the \textit{condemned set}. 
The \textbf{black} set is the set of objects that have no references to objects in the white set, and are unreachable from the root. The \textbf{grey} set is the set of objects reachable from the root but whose references have not yet been scanned.

For the entire heap, including the roots, every object is assigned a color thereby partitioning the memory into the three sets. For the most part, algorithms implementing some variation of the tri-color abstraction execute some combination of the following instructions: 
\begin{enumerate}
  \item pick an object from the gray set
  \item gray all of the objects references and move it to the black set
  \item repeat steps 1 and 2 until the gray set is empty.
  \item we define the remaining objects as black if they are reachable from the root and white if they are unreachable
  \item collect preform garbage collection on the white objects
\end{enumerate}

\subsection{Mark and Sweep}
The earliest and most basic garbage collection algorithm is Mark-Sweep garbage collection\cite{mccarthy} - or a copy collector\footnote{we will use these terms interchangeable throughout the paper}, and most modern algorithms are a variant on it. Mark-Sweep is a “stop-the world” collector, which means that at some point when the program requests memory and none is available, the program is stopped and a full garbage collection is performed to free up space. 

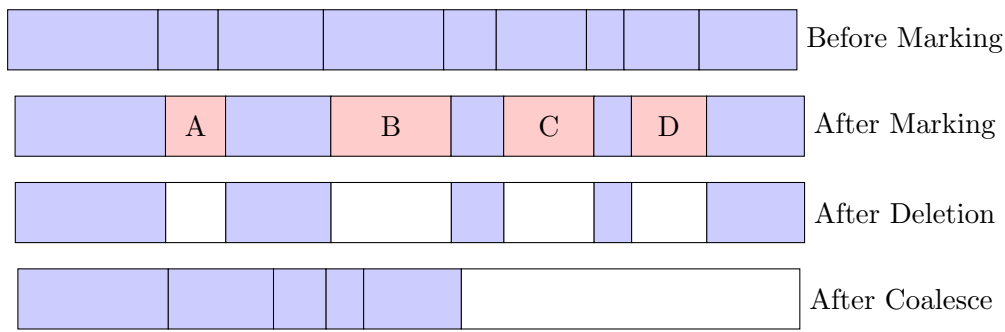
\begin{figure}[H]
	\centering
      \begin{tikzpicture}
          [start chain=1 going right,node distance=-0.15mm]
          \node (rect) [draw,on chain=1, style={fill=blue!20},minimum width=20mm,minimum height=8mm] {};
          \node (rect) [draw,on chain=1, style={fill=blue!20},minimum width=8mm,minimum height=8mm] {};
          \node (rect) [draw,on chain=1, style={fill=blue!20},minimum width=14mm,minimum height=8mm] {};
          \node (rect) [draw,on chain=1, style={fill=blue!20},minimum width=16mm,minimum height=8mm] {};
          \node (rect) [draw,on chain=1, style={fill=blue!20},minimum width=7mm,minimum height=8mm] {};
          \node (rect) [draw,on chain=1, style={fill=blue!20},minimum width=12mm,minimum height=8mm] {};
          \node (rect) [draw,on chain=1, style={fill=blue!20},minimum width=5mm,minimum height=8mm] {};
          \node (rect) [draw,on chain=1, style={fill=blue!20},minimum width=10mm,minimum height=8mm] {};
          \node (rect) [draw,on chain=1, style={fill=blue!20},minimum width=13mm,minimum height=8mm] {};
          \node (rect) [name=r,on chain=1] {Before Marking};
      \end{tikzpicture}\\
	  \vspace{3mm}
	  \begin{tikzpicture}
          [start chain=1 going right,node distance=-0.15mm]
          \node (rect) [draw,on chain=1, style={fill=blue!20},minimum width=20mm,minimum height=8mm] {};
          \node (rect) [draw,on chain=1, style={fill=red!20},minimum width=8mm,minimum height=8mm] {A};
          \node (rect) [draw,on chain=1, style={fill=blue!20},minimum width=14mm,minimum height=8mm] {};
          \node (rect) [draw,on chain=1, style={fill=red!20},minimum width=16mm,minimum height=8mm] {B};
          \node (rect) [draw,on chain=1, style={fill=blue!20},minimum width=7mm,minimum height=8mm] {};
          \node (rect) [draw,on chain=1, style={fill=red!20},minimum width=12mm,minimum height=8mm] {C};
          \node (rect) [draw,on chain=1, style={fill=blue!20},minimum width=5mm,minimum height=8mm] {};
          \node (rect) [draw,on chain=1, style={fill=red!20},minimum width=10mm,minimum height=8mm] {D};
          \node (rect) [draw,on chain=1, style={fill=blue!20},minimum width=13mm,minimum height=8mm] {};
          \node (rect) [name=r,on chain=1] {After Marking};
      \end{tikzpicture}\\
	  \vspace{3mm}
	  \begin{tikzpicture}
          [start chain=1 going right,node distance=-0.15mm]
          \node (rect) [draw,on chain=1, style={fill=blue!20},minimum width=20mm,minimum height=8mm] {};
          \node (rect) [draw,on chain=1, style={fill=white!20},minimum width=8mm,minimum height=8mm] {};
          \node (rect) [draw,on chain=1, style={fill=blue!20},minimum width=14mm,minimum height=8mm] {};
          \node (rect) [draw,on chain=1, style={fill=white!20},minimum width=16mm,minimum height=8mm] {};
          \node (rect) [draw,on chain=1, style={fill=blue!20},minimum width=7mm,minimum height=8mm] {};
          \node (rect) [draw,on chain=1, style={fill=white!20},minimum width=12mm,minimum height=8mm] {};
          \node (rect) [draw,on chain=1, style={fill=blue!20},minimum width=5mm,minimum height=8mm] {};
          \node (rect) [draw,on chain=1, style={fill=white!20},minimum width=10mm,minimum height=8mm] {};
          \node (rect) [draw,on chain=1, style={fill=blue!20},minimum width=13mm,minimum height=8mm] {};
          \node (rect) [name=r,on chain=1] {After Deletion};
      \end{tikzpicture}\\
      \vspace{3mm}
	  \begin{tikzpicture}
          [start chain=1 going right,node distance=-0.15mm]
          \node (rect) [draw,on chain=1, style={fill=blue!20},minimum width=20mm,minimum height=8mm] {};
          \node (rect) [draw,on chain=1, style={fill=blue!20},minimum width=14mm,minimum height=8mm] {};
          \node (rect) [draw,on chain=1, style={fill=blue!20},minimum width=7mm,minimum height=8mm] {};
          \node (rect) [draw,on chain=1, style={fill=blue!20},minimum width=5mm,minimum height=8mm] {};
          \node (rect) [draw,on chain=1, style={fill=blue!20},minimum width=13mm,minimum height=8mm] {};
          \node (rect) [draw,on chain=1, style={fill=white!20},minimum width=45mm,minimum height=8mm] {};
          \node (rect) [name=r,on chain=1] {After Coalesce};
      \end{tikzpicture}\\
	\caption{Mark and Sweep Process}
	\label{fig:marksweep}
\end{figure}

As detailed in Figure~\ref{fig:heapex} above, in the Mark-Sweep algorithm, each object has a “mark-bit” which is used during the collection process to track whether the object has been visited. Here is an algorithm for Mark-Sweep garbage collection implemented on top of some underlying explicit memory management routines, in which free regions of the heap are also considered objects with mark bits and a known size. According to McCarthy\cite{mccarthy}, the ``stop-the-world'' label was an effective description of the process involved with this garbage collector. The efficiency of a GC system depends primarily upon not reaching the exhaustion point of available memory - the reason being that the memory reclamation process, according to McCarthy, required several seconds and the addition of several thousand registers to the free-list.

\subsection{Cheney's Copy Algorithm}
\begin{figure}[H]
	\centering
      \begin{tikzpicture}
          [start chain=1 going right,node distance=-0.15mm]
          \node (rect) [draw,on chain=1, style={fill=white!20},minimum width=70mm,minimum height=8mm] {};
          \node (rect) [draw,on chain=1, style={fill=blue!20},minimum width=12mm,minimum height=8mm] {};
          \node (rect) [draw,on chain=1, style={fill=red!20},minimum width=8mm,minimum height=8mm] {C};
          \node (rect) [draw,on chain=1, style={fill=red!20},minimum width=14mm,minimum height=8mm] {A};
          \node (rect) [draw,on chain=1, style={fill=blue!20},minimum width=16mm,minimum height=8mm] {};
          \node (rect) [draw,on chain=1, style={fill=red!20},minimum width=7mm,minimum height=8mm] {D};
          \node (rect) [draw,on chain=1, style={fill=red!20},minimum width=12mm,minimum height=8mm] {B};
          \node at (0.0,0.7) {To-Space};
          \node at (7.0,0.7) {From-Space};
          \node (rect) [name=r,on chain=1] {Before Collection};
      \end{tikzpicture}\\
      
       \begin{tikzpicture}
          [start chain=1 going right,node distance=-0.15mm]
          \node (rect) [draw,on chain=1, style={fill=red!20},minimum width=14mm,minimum height=8mm] (a1) {A'};
          \node (rect) [draw,on chain=1, style={fill=red!20},minimum width=12mm,minimum height=8mm] (b1) {B'};
          \node (rect) [draw,on chain=1, style={fill=red!20},minimum width=8mm,minimum height=8mm] (c1) {C'};
          \node (rect) [draw,on chain=1, style={fill=red!20},minimum width=7mm,minimum height=8mm] (d1) {D'};
          \node (rect) [draw,on chain=1, style={fill=white!20},minimum width=29mm,minimum height=8mm] {};

          \node (rect) [draw,on chain=1, style={fill=blue!20},minimum width=12mm,minimum height=8mm] {};
          \node (rect) [draw,on chain=1, style={fill=red!20},minimum width=8mm,minimum height=8mm] (c) {C};
          \node (rect) [draw,on chain=1, style={fill=red!20},minimum width=14mm,minimum height=8mm] (a) {A};
          \node (rect) [draw,on chain=1, style={fill=blue!20},minimum width=16mm,minimum height=8mm] {};
          \node (rect) [draw,on chain=1, style={fill=red!20},minimum width=7mm,minimum height=8mm] (d) {D};
          \node (rect) [draw,on chain=1, style={fill=red!20},minimum width=12mm,minimum height=8mm] (b) {B};
         
          \draw[->] [out=-155, in=-30] (a) to node[below]{} (a1);
          \draw[->] [out=-155, in=-30] (b) to node[below]{} (b1);
          \draw[->] [out=-155, in=-30] (c) to node[below]{} (c1);
          \draw[->] [out=-155, in=-30] (d) to node[below]{} (d1);
          
          \node at (0.0,0.7) {To-Space};
          \node at (7.0,0.7) {From-Space};
          \node (rect) [name=r,on chain=1] {After Copying};
      \end{tikzpicture}\\
      
      \begin{tikzpicture}
          [start chain=1 going right,node distance=-0.15mm]
          \node (rect) [draw,on chain=1, style={fill=red!20},minimum width=14mm,minimum height=8mm] {A'};
          \node (rect) [draw,on chain=1, style={fill=red!20},minimum width=12mm,minimum height=8mm] {B'};
          \node (rect) [draw,on chain=1, style={fill=red!20},minimum width=8mm,minimum height=8mm] {C'};
          \node (rect) [draw,on chain=1, style={fill=red!20},minimum width=7mm,minimum height=8mm] {D'};
          \node (rect) [draw,on chain=1, style={fill=white!20},minimum width=29mm,minimum height=8mm] {};
          \node (rect) [draw,on chain=1, style={fill=blue!20},minimum width=69mm,minimum height=8mm] {};
          \node at (3,0.7) {To-Space};
          \node at (10.0,0.7) {From-Space};
          \node (rect) [name=r,on chain=1] {After Copying};
      \end{tikzpicture}\\

	\caption{Cheney's Copy Process}
	\label{fig:cheney}
\end{figure}
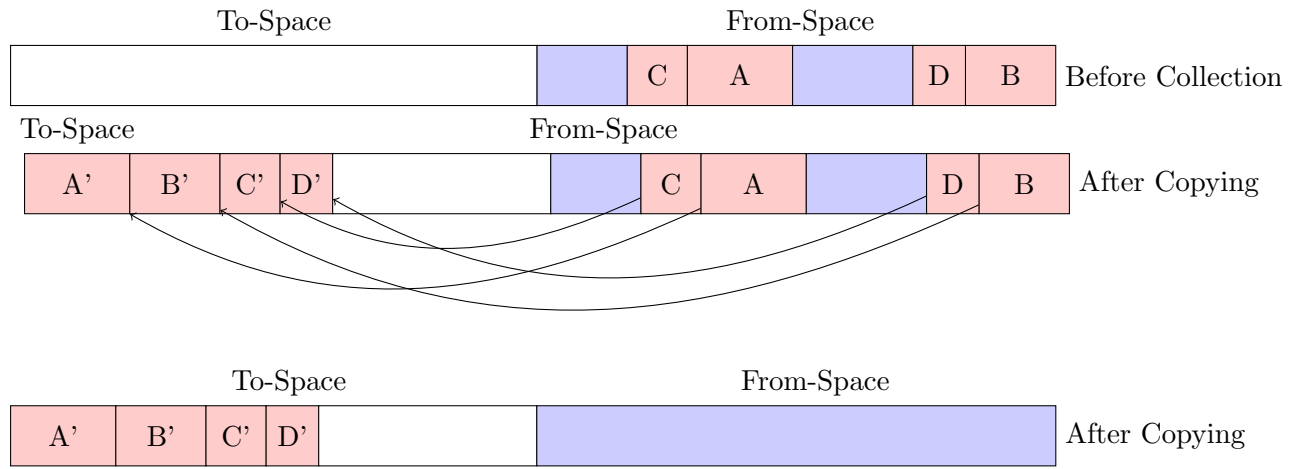

Cheney's Algorithm is a subset of garbage collectors known as copying collectors\cite{wiki:garbage_collection}. In copying collectors, reachable objects are relocated from one address to another during a collection. Available memory is divided into two equal-size regions called the \mydef{from-space} and the \mydef{to-space}. During allocation, the system keeps a pointer into the to-space which is incremented by the amount of memory requested for each allocation. When there is insufficient space in to-space to fulfill an allocation, a collection is performed. A collection consists of swapping the roles of the regions, and copying the live objects from from-space to to-space, leaving a block of free space (corresponding to the memory used by all unreachable objects) at the end of the to-space. Since objects are moved during a collection, the addresses of all references must be updated. This is done by storing a “forwarding-address” for an object when it is copied out of from-space. Like the mark-bit, this forwarding-address can be thought of as an additional field of the object, but is usually implemented by temporarily repurposing some space from the object. 

\subsection{Knuth's Lisp 2 Algorithm}
\usetikzlibrary{shapes,arrows}

\tikzstyle{block} = [draw, fill=blue!20, rectangle, 
    minimum height=3em, minimum width=6em]
\tikzstyle{sum} = [draw, fill=blue!20, circle, node distance=1cm]
\tikzstyle{input} = [coordinate]
\tikzstyle{output} = [coordinate]
\tikzstyle{pinstyle} = [pin edge={to-,thin,black}]

\begin{figure}[H]
	\centering   
       \begin{tikzpicture}
           [start chain=1 going right,node distance=-0.15mm]
          \node (rect) [draw,on chain=1, style={fill=white!20},minimum width=2mm,minimum height=8mm] (r) {r};
          \node (rect) [draw,on chain=1, style={fill=white!20},minimum width=20mm,minimum height=8mm] (n1) {start};
          \node (rect) [draw,on chain=1, style={fill=blue!20},minimum width=8mm,minimum height=8mm] (n2) {A};
          \node (rect) [draw,on chain=1, style={fill=white!20},minimum width=14mm,minimum height=8mm] (n3) {};
          \node (rect) [draw,on chain=1, style={fill=blue!20},minimum width=16mm,minimum height=8mm] (n4) {B};
          \node (rect) [draw,on chain=1, style={fill=white!20},minimum width=7mm,minimum height=8mm] (n5) {};
          \node (rect) [draw,on chain=1, style={fill=blue!20},minimum width=12mm,minimum height=8mm] (n6) {C};
          \node (rect) [draw,on chain=1, style={fill=white!20},minimum width=5mm,minimum height=8mm] (n7) {};
          \node (rect) [draw,on chain=1, style={fill=blue!20},minimum width=10mm,minimum height=8mm] (n8) {D};
          \node (rect) [draw,on chain=1, style={fill=white!20},minimum width=13mm,minimum height=8mm] (n9) {};
          \node (rect) [draw,on chain=1, style={fill=blue!20},minimum width=10mm,minimum height=8mm] (n10) {E};
          \node (rect) [draw,on chain=1, style={fill=white!20},minimum width=13mm,minimum height=8mm] (n11) {end};
		   \draw[out=45, in=135] (r) to node[above]{} (n6);
		   \draw[out=45, in=135] (r) to node[above]{} (n8);
           \draw[out=45, in=135] (r) to node[above]{} (n10);
           \draw[out=45, in=135] (n2) to node[above]{} (n10);
           \draw[out=45, in=135] (n4) to node[above]{} (n8);
           \draw[out=45, in=135] (n6) to node[above]{} (n6);

        \node (rect) [name=r,on chain=1] {Init. Structure};
      \end{tikzpicture}

      \begin{tikzpicture}
           [start chain=1 going right,node distance=-0.15mm]
          \node (rect) [draw,on chain=1, style={fill=white!20},minimum width=2mm,minimum height=8mm] (r) {r};
          \node (rect) [draw,on chain=1, style={fill=white!20},minimum width=20mm,minimum height=8mm] (n1) {start};
          \node (rect) [draw,on chain=1, style={fill=blue!20},minimum width=8mm,minimum height=8mm] (n2) {A};
          \node (rect) [draw,on chain=1, style={fill=white!20},minimum width=14mm,minimum height=8mm] (n3) {};
          \node (rect) [draw,on chain=1, style={fill=blue!20},minimum width=16mm,minimum height=8mm] (n4) {B};
          \node (rect) [draw,on chain=1, style={fill=white!20},minimum width=7mm,minimum height=8mm] (n5) {};
          \node (rect) [draw,on chain=1, style={fill=blue!20},minimum width=12mm,minimum height=8mm] (n6) {C};
          \node (rect) [draw,on chain=1, style={fill=white!20},minimum width=5mm,minimum height=8mm] (n7) {};
          \node (rect) [draw,on chain=1, style={fill=blue!20},minimum width=10mm,minimum height=8mm] (n8) {D};
          \node (rect) [draw,on chain=1, style={fill=white!20},minimum width=13mm,minimum height=8mm] (n9) {};
          \node (rect) [draw,on chain=1, style={fill=blue!20},minimum width=10mm,minimum height=8mm] (n10) {E};
          \node (rect) [draw,on chain=1, style={fill=white!20},minimum width=13mm,minimum height=8mm] (n11) {end};
		   \draw[out=45, in=135] (r) to node[above]{} (n6);
		   \draw[out=45, in=135] (r) to node[above]{} (n8);
           \draw[out=45, in=135] (r) to node[above]{} (n10);
           \draw[out=45, in=135] (n2) to node[above]{} (n10);
           \draw[out=45, in=135] (n4) to node[above]{} (n8);
           \draw[out=45, in=135] (n6) to node[above]{} (n6);
           
           \draw[->] [out=-135, in=-45] (n2) to node[below]{} (n1);
           \draw[->] [out=-135, in=-45] (n4) to node[below]{} (n2);
           \draw[->] [out=-135, in=-45] (n6) to node[below]{} (n3);
           \draw[->] [out=-135, in=-45] (n8) to node[below]{} (n3);
           \draw[->] [out=-135, in=-45] (n10) to node[below]{}(n5);         
           
        \node (rect) [name=r,on chain=1] {After Pass 1};
      \end{tikzpicture}
      
      \begin{tikzpicture}
           [start chain=1 going right,node distance=-0.15mm]
          \node (rect) [draw,on chain=1, style={fill=white!20},minimum width=2mm,minimum height=8mm] (r) {r};
          \node (rect) [draw,on chain=1, style={fill=white!20},minimum width=20mm,minimum height=8mm] (n1) {start};
          \node (rect) [draw,on chain=1, style={fill=blue!20},minimum width=8mm,minimum height=8mm] (n2) {A};
          \node (rect) [draw,on chain=1, style={fill=white!20},minimum width=14mm,minimum height=8mm] (n3) {};
          \node (rect) [draw,on chain=1, style={fill=blue!20},minimum width=16mm,minimum height=8mm] (n4) {B};
          \node (rect) [draw,on chain=1, style={fill=white!20},minimum width=7mm,minimum height=8mm] (n5) {};
          \node (rect) [draw,on chain=1, style={fill=blue!20},minimum width=12mm,minimum height=8mm] (n6) {C};
          \node (rect) [draw,on chain=1, style={fill=white!20},minimum width=5mm,minimum height=8mm] (n7) {};
          \node (rect) [draw,on chain=1, style={fill=blue!20},minimum width=10mm,minimum height=8mm] (n8) {D};
          \node (rect) [draw,on chain=1, style={fill=white!20},minimum width=13mm,minimum height=8mm] (n9) {};
          \node (rect) [draw,on chain=1, style={fill=blue!20},minimum width=10mm,minimum height=8mm] (n10) {E};
          \node (rect) [draw,on chain=1, style={fill=white!20},minimum width=13mm,minimum height=8mm] (n11) {end};
		   \draw[out=45, in=135] (r) to node[above]{} (n2);
           \draw[->] [out=45, in=135] (n2) to node[above]{} (n3);
           \draw[out=45, in=135] (r) to node[above]{} (n4);
		   \draw[out=45, in=135] (r) to node[above]{} (n6);
		   \draw[out=45, in=135] (r) to node[above]{} (n8);
           \draw[out=45, in=135] (r) to node[above]{} (n10);
           
           \draw[->] [dashed, out=-135, in=-45] (n4) to node[below]{} (n2);
           \draw[->] [dashed, out=-135, in=-45] (n6) to node[below]{} (n3);
           \draw[->] [dashed, out=-135, in=-45] (n8) to node[below]{} (n3);
           \draw[->] [dashed, out=-135, in=-45] (n10) to node[below]{}(n5);
           
        \node (rect) [name=r,on chain=1] {After Pass 2};
      \end{tikzpicture}
            
      \begin{tikzpicture}
           [start chain=1 going right,node distance=-0.15mm]
          \node (rect) [draw,on chain=1, style={fill=white!20},minimum width=2mm,minimum height=8mm] (r) {r};
          \node (rect) [draw,on chain=1, style={fill=white!20},minimum width=20mm,minimum height=8mm] (n1) {start};
          \node (rect) [draw,on chain=1, style={fill=blue!20},minimum width=8mm,minimum height=8mm] (n2) {A};
          \node (rect) [draw,on chain=1, style={fill=blue!20},minimum width=16mm,minimum height=8mm] (n4) {B};
          \node (rect) [draw,on chain=1, style={fill=blue!20},minimum width=12mm,minimum height=8mm] (n6) {C};
          \node (rect) [draw,on chain=1, style={fill=blue!20},minimum width=10mm,minimum height=8mm] (n8) {D};
          \node (rect) [draw,on chain=1, style={fill=blue!20},minimum width=10mm,minimum height=8mm] (n10) {end(E)};
          \node (rect) [draw,on chain=1, style={fill=white!20},minimum width=52mm,minimum height=8mm] (n3) {};
          \
          
		   \draw[out=45, in=135] (n2) to node[above]{} (n8);
           \draw[out=45, in=135] (n4) to node[above]{} (n10);
		   \draw[out=45, in=135] (r) to node[above]{} (n6);
		   \draw[out=45, in=135] (r) to node[above]{} (n8);
           \draw[out=45, in=135] (r) to node[above]{} (n10);
                
        \node (rect) [name=r,on chain=1] {After Pass 3};
      \end{tikzpicture}
      
	\caption{Lisp 2 Process}
	\label{fig:lisp2}
\end{figure}
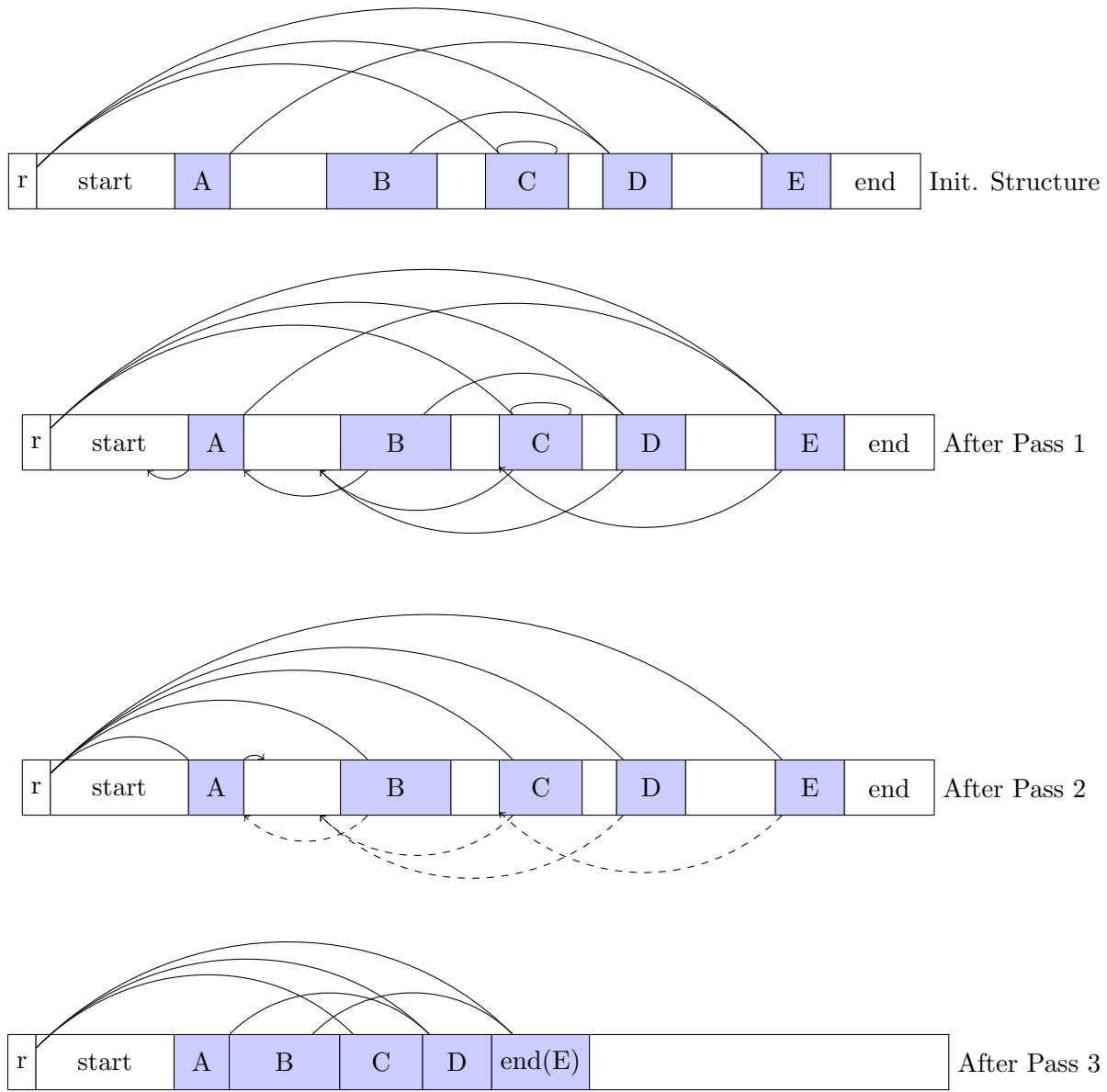

Knuth's Lisp 2 Algorithm\cite{knuth3} is a classical algorithm developed for a lisp-based language that involves
frequent allocation and deallocation of many small objects. Detailed by Knuth, the Lisp 2 Algorithm is a ``compacting algorithm'' which utilizes three linear passes of the memory space to compute and optimize pointers to objects in memory. In the first pass, the algorithm computes a new address for each active cell. During the second pass, the algorithm updates pointers to each active cells. The third and final pass is composed of reallocation of active cells and moving objects such that active, allocated objects are grouped together.

\section{Analysis and Discussion}
We analyze each algorithm with respect to asymptotic time complexity and determine that all algorithms should yield an asymptotic complexity of linear $O(n)$ time and space with respect to the the size of the heap, with varying constants.

\subsection{Complexity}
We assert asymptotic behavior is not an effective metric when used to compare the efficiency of garbage collection algorithms. To simplify our analysis, we primarily compare the complexity of the Mark-Sweep and Copy Collector garbage collectors defined in section 3. We generalize the phases of the Mark-Sweep again below:
\begin{enumerate}
  \item Initialize traces by clearing marked bits
  \item Trace
  \item Mark
  \item Sweep
  \item Reclaim allocated memory
\end{enumerate}
We can say that phase 1 has a negligible effect on memory and time for both garbage collectors, thus we examine the complexity of both algorithm's with respect to phases 2, 3, 4, and 5.

When examining Cheney's Copy Collector, it is important to note that the algorithm forgoes direct allocation and instead swaps objects in regions of memory\cite{hertz}. Phase 3 of the Mark-Sweep algorithm can also be disregarded in the context of a Copy Collector. The sweep phase is linearly proportional to the number of active objects in the heap for both collectors.
Regarding phase 4, since Cheney's algorithm in particular passes through the heap breadth first, we can say the complexity is linearly proportional to the number of active cells - a potential significant difference to the total size of the heap. In phase 5, we say that the time complexity is proportional to the size of the uncopied objects in the heap. Thus, we can conclude that asymptotic complexity of preforming collection on a heap of n objects is proportional to n regardless of the algorithm used.

According to Hertz, despite having worse worst-case space performance, for realistic applications, the Mark-Sweep algorithm is preferable in terms of space, but the copying collection offers much faster allocation than the traditional sweep and allocate technique used by Mark-Sweep.\cite{hertz}

Despite stating the equivalence of asymptotic complexity, we can conclude that for most practical applications involving a lisp-like language Cheney's Copy Collector surpasses a traditional Mark-Sweep GC in terms of speed and worst-case space performance.

\subsection{A note about generality}
As alluded to earlier, the efficiency of garbage collectors is a difficult aspect to quantify\cite{hertz}. Furthermore, it is evident that particular garbage collectors, while effective when programming under certain languages or paradigms, may not be effective under others. Just by recognizing the sheer number of garbage collectors in use today and the varying efficiency of those languages, it becomes apparent that developing an effective general case garbage collector is not a feasible task.

\section{Results}
\subsection{Testing}
During development, it is vital to have some way to examine the state of the heap during run time and identify anomalous behavior. Freeing nodes which are still in use can be identified rather easily, as this will trigger a segmentation fault when trying to access freed data. More difficult, is identifying unused data which is not freed, which will cause no external problems other than consuming more memory than necessary. To make sure that all nodes are being properly allocated and freed during run time, we take advantage of \emph{Valgrind}, a dynamic analysis program for arbitrary binaries. Valgrind is a suite of tools which can be used to identify a variety of subtle errors that can present during run time, including memory leaks and improper memory access. As a part of testing, we use Valgrind to verify the correctness of our allocation system, and to help locate the source of errors.

\subsection{Implementation}
Our project was designed from the beginning to use a loose coupling between the parser, evaluator, REPL etc. and the garbage collector. That way, we can simply ``drag and drop'' different garbage collection strategies without having to modify any other code. In this manner, we implemented the core interpreter and parser, along with the Cheney's algorithm and Mark-Sweep with Tri-color marking. This allows us to evaluate both strategies on equal footing.

For comparing the two strategies, we consider two important metrics: average allocation time and average GC time. Allocation time refers to how long it takes to create a new object on the heap, including a possible garbage collection pause. GC time refers to the amount of time it takes for the actual garbage collection algorithm to run. These statistics are gathered during run time by using the high precision timers in the C++ \emph{chrono} library, which allows the collection of timing data accurate on the order of nanoseconds. Testing itself was performed on the Major's lab computers. These machines feature an 8-core Intel i7 processor with 8 MB of Cache and 8 GB of RAM.

The actual lisp script used is described in the appendix and includes a variety of inputs to test correctness and also demonstrate performance. This includes several recursive functions and the dynamic generation of lists, as well as the usage of a user-defined higher-order function \emph{map}. Most intensive is the final line: \emph{(map (LAMBDA (x) (fib x)) (range 0 20))}, which generates a list of the first 20 fibonacci numbers, using the higher order map function, and a naive recursive implementation of the fibonacci function. Evaluating this script requires the allocation of over a quarter of a million nodes, which can require anywhere from $400$ to $1800$ garbage collections, depending on the size of the heap. To put additional pressure on the garbage collectors, the heap is artificially constrained to between 8 and 15 kilobytes.

\subsection{Allocation Timing}

Cheney's algorithm has a key advantage over non-copying techniques for allocation timing because it does not require the use of a complete allocator, with freeing and coalescing. Because the entire heap is copied over and freed all at once, it simply needs to use a ``bump pointer'' for allocation: the algorithm simply tracks the end of the allocated heap and increments this pointer with each new allocation. As a result, Cheney's algorithm works faster than Mark-Sweep by a factor of four for allocation. In all cases tested, average allocation time remained less than 100 nanoseconds for Cheney's algorithm. 

A second important observation is that the average allocation time is inversely proportional to the size of the heap. This is because a larger heap will require garbage collection pauses to occur less often. This phenomenon holds true regardless of GC strategy.

\subsection{Garbage Collection Timing}

Cheney's algorithm also significantly outperformed Mark-Sweep for GC pause time, and also proved to be far more stable with respect to heap size. For Mark-Sweep, GC pause time is clearly proportional to heap size, but that effect is less clear for Cheney's algorithm. Cheney's algorithm remains relatively constant regardless of heap size. In all cases tested, the average GC pause time for Cheney's algorithm remained under 30 thousand nanoseconds. Even when the heap size was extended to 1 megabyte in size, and order of magnitude larger than other tested values, the average GC pause time averaged only 50 thousand nanoseconds. In comparison, Mark-Sweep run with a 1 megabyte heap averages 4.9 \emph{million} nanosecond GC pauses: three orders of magnitude longer than Cheney's algorithm. While not rigorous, this is strong evidence showing that the average performance of Cheney's algorithm is actually sublinear with respect to the heap size. This is largely due to the fact that Cheney's algorithm requires examining only live nodes, while Mark-Sweep must scan through the entire heap and remove dead nodes during the Sweep phase. For our lisp script, the majority of nodes have very short lifetimes, meaning the majority of the heap is filled with dead nodes.

\section{Concluding Statements}
Since a Mark-Sweep collector traverses the heap by following pointers to referenced objects starting at the roots, it is necessary for a Mark-Sweep collector to traverse the heap multiple times in order to mark and sweep objects to be freed. In contrast, a Copy collector copies objects reachable from a root note from the from-space in memory to the to-space as the nodes are traversed with Cheney's bread-first search. As we have stated in this paper, both garbage collectors can be utilized effectively in different environments, or for different languages.

\section{Future Work and Open Problems}
Garbage collectors are far from perfect. As a testament to the overhead required by their use - the toll they take on the speed and efficiency of a program with which they are used, they are rarely used in embedded or real-time systems where tight control over limited resources is essential. Continued research in platforms compatible with lightweight and fast programs could be conducted under avenues such as multithreading, Operating System independence or MMU independence. 

\section*{Acknowledgments}
We thank Professor Lane A. Hemaspaandra and Teaching Assistants Charles Lehner, Taylan Sen, and Adam Scrivener for providing helpful direction and an environment that encouraged us to collaborate and improve our collective understanding of Lisp, the structure of memory in the heapspace, and garbage collecting algorithms. 

\nocite{*}
\printbibliography

\section*{Appendix}
\textbf{Sample Output}
\begin{lstlisting}[breaklines]
minilisp> (QUOTE A)
A
minilisp> (QUOTE (A B C))
(A B C)
minilisp> (CAR (QUOTE (A B C)))
A
minilisp> (CDR (QUOTE (A B C)))
(B C)
minilisp> (CONS (QUOTE A) (QUOTE (B C)))
(A B C)
minilisp> (= (CAR (QUOTE (A B))) (QUOTE A))
#T
minilisp> (= (CAR (CDR (QUOTE (A B)))) (QUOTE A))
#F
minilisp> (CAR (QUOTE (0 1)))
0
minilisp> (CDR (CONS (+ 0 1) (+ 2 3)))
5
minilisp> (DEFINE foo (+ 0 1))
NIL
minilisp> foo
1
minilisp> (DEFINE bar 0)
NIL
minilisp> bar
0
minilisp> (+ foo bar)
1
minilisp> (COND (#T (+ 0 1)))
1
minilisp> (COND ((= foo bar) 7) (#T 9))
9
minilisp> (COND ((= (QUOTE A) (QUOTE B)) (QUOTE C)) ((NOT #F) (QUOTE yee)))
yee
minilisp> ((LAMBDA (X) (+ X 1)) 5)
6
minilisp> (DEFINE add (LAMBDA (X) (LAMBDA (Y) (+ X Y))))
NIL
minilisp> ((add 4) 5)
9
minilisp> (DEFINE fac (LAMBDA (N) (COND ((= N 0) 1) (#T (* N (fac (- N 1)))))))
NIL
minilisp> (fac 0)
1
minilisp> (fac 10)
3628800
minilisp> (DEFINE range (LAMBDA (LOW HIGH) (COND ((> LOW HIGH) NIL) (#T (CONS LOW (range (+ LOW 1) HIGH))))))
NIL
minilisp> (range 0 100)
(0 1 2 3 4 5 6 7 8 9 10 11 12 13 14 15 16 17 18 19 20 21 22 23 24 25 26 27 28 29 30 31 32 33 34 35 36 37 38 39 40 41 42 43 44 45 46 47 48 49 50 51 52 53 54 55 56 57 58 59 60 61 62 63 64 65 66 67 68 69 70 71 72 73 74 75 76 77 78 79 80 81 82 83 84 85 86 87 88 89 90 91 92 93 94 95 96 97 98 99 100)
minilisp> (DEFINE map (LAMBDA (f xs) (COND ((= xs NIL) NIL) (#T (CONS (f (CAR xs)) (map f (CDR xs)))))))
NIL
minilisp> (map (LAMBDA (x) (+ x 1)) (range 0 50))
(1 2 3 4 5 6 7 8 9 10 11 12 13 14 15 16 17 18 19 20 21 22 23 24 25 26 27 28 29 30 31 32 33 34 35 36 37 38 39 40 41 42 43 44 45 46 47 48 49 50 51)
minilisp> (map (LAMBDA (x) (fac x)) (range 0 15))
(1 1 2 6 24 120 720 5040 40320 362880 3628800 39916800 479001600 1932053504 1278945280 2004310016)
minilisp> (DEFINE fib (LAMBDA (n) (COND ((OR (= n 0) (= n 1)) 1) (#T (+ (fib (- n 1)) (fib (- n 2)))))))
NIL
minilisp> (map (LAMBDA (x) (fib x)) (range 0 20))
(1 1 2 3 5 8 13 21 34 55 89 144 233 377 610 987 1597 2584 4181 6765 10946)
ALLOC TIME: 158859467
TOTAL ALLOCATIONS: 259529
AVG ALLOCATION TIME: 612
GC TIME: 121797803
TOTAL GC COLLECTIONS: 959
AVG GC TIME: 127005
\end{lstlisting}

\textbf{Graphs}
\begin{figure}[H]
	\centering
    	\includegraphics[width=8cm,scale=.3]{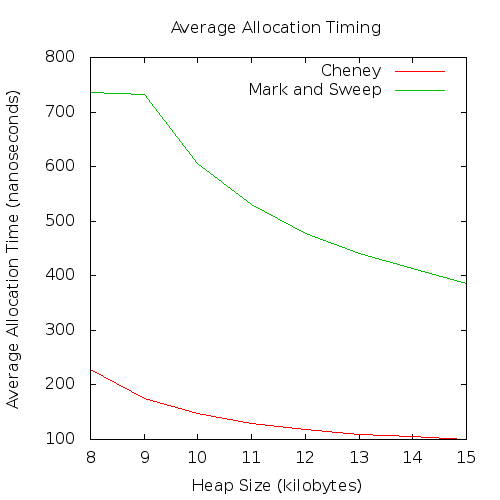}
	\caption{Average Allocation Time vs Heap Size}
	\label{fig:alloct}
\end{figure}

\begin{figure}[H]
	\centering
    	\includegraphics[width=8cm,scale=.3]{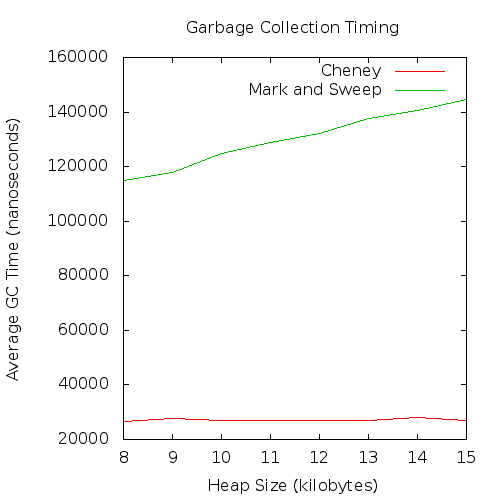}
	\caption{Average Collection Time vs Heap Size}
	\label{fig:gct}
\end{figure}

\end{document}